\documentclass[a4paper,11pt]{article}
\usepackage{jinstpub} % for details on the use of the package, please see the JINST-author-manual
\usepackage{lineno}
\usepackage{graphicx}% Include figure files
\usepackage{dcolumn}% Align table columns on decimal point
\usepackage{bm}% bold math
\usepackage{color}
\usepackage[dvipsnames]{xcolor}
\usepackage[utf8]{inputenc}
\usepackage[T1]{fontenc}
\usepackage{newtxtext,newtxmath}
\usepackage[dvipsnames]{xcolor}
\usepackage{multirow}
\usepackage{threeparttable}
\usepackage{booktabs}
\usepackage{float}
\usepackage{amsmath}
\usepackage{gensymb}
\usepackage{caption}
\usepackage{subcaption}

\everymath{\displaystyle}

\usepackage{enumitem}
\setlist[enumerate]{label*=\arabic*.}

\usepackage{xcolor}
\def\ow#1{{\textcolor{red}{[OW: {#1}]}}}

\newcommand{\bigo}[1]{$\mathcal{O}$(#1)}

\def\tqp{\tau_{\rm qp}}

\title{Significant noise improvement in a Kinetic Inductance Phonon-Mediated detector by use of a wideband parametric amplifier}

\author[1]{K.~Ramanathan}
\author[2,6]{O.~Wen}
\author[2,6]{T.~Aralis}
\author[3]{R.~Basu~Thakur}
\author[3]{B.~Bumble}
\author[4]{Y.-Y.~Chang}
\author[3]{P.~K.~Day}
\author[3]{B.~H.~Eom}
\author[3]{H.~G.~Leduc}
\author[5]{B.~J.~Sandoval}
\author[5]{R.~Stephenson}
\author[5]{S.~R.~Golwala}

\affiliation[1]{Department of Physics, Washington University in St. Louis, 1 Brookings Dr, St. Louis, MO 63130, U.S.A}
\affiliation[2]{SLAC National Accelerator Laboratory, 2575 Sand Hill Rd, Menlo Park, CA 94025, U.S.A}
\affiliation[3]{Jet Propulsion Laboratory, California Institute of Technology, 4800 Oak Grove Dr, Pasadena, CA 91109, U.S.A}
\affiliation[4]{Department of Physics, University of California, Berkeley, 366 LeConte Hall, Berkeley, CA 94720, U.S.A.}
\affiliation[5]{Department of Physics, California Institute of Technology, 1200 E California Blvd, Pasadena, CA 91125, U.S.A}
\affiliation[6]{Kavli Institute for Particle Astrophysics and Cosmology, Stanford University, 450 Jane Stanford Way, Stanford, CA 94305, U.S.A.}

\emailAdd{karthikr@wustl.edu}
\emailAdd{osmondw@wustl.edu}


\abstract{
Microwave Kinetic Inductance Detectors (MKIDs) have been demonstrated as capable phonon sensors when coupled to crystalline substrates, and have been proposed as detectors for next-generation rare-event searches such as for the direct detection of dark matter. These Kinetic Inductance Phonon-Mediated (KIPM) detector designs, favoring large superconducting absorber volumes and high readout powers, are oftentimes limited in their sensitivity by low temperature amplifier noise introduced in the signal readout chain. We report here an effort to couple a wideband Kinetic Inductance Traveling Wave Parametric Amplifier (KI-TWPA), operated near the Standard Quantum Limit of minimal added amplifier noise, to sensors spanning a 70 MHz bandwidth at 3.5~GHz. This results in a $\sim$5$\times$ improvement in the inferred detector energy resolution in the best sensor and highlights the potential of constructing $\mathcal{O}$(100)~meV resolving phonon-mediated particle detectors. We detail limitations introduced by lossy passive components, degraded RF responsivity, and microphysical noise sources like two-level systems (TLS), in achieving ultimate quantum-limited system noise levels.
}

\begin{document}
\maketitle
\flushbottom

\section{Introduction and Motivation}

Microwave Kinetic Inductance Detectors (MKIDs) are thin film superconducting LC microresonators that originated as photon sensors \cite{day2003}. The resonator surface impedance contains a reactive component due to Cooper-pair inertia, termed the \textit{kinetic inductance}. If an MKID absorbs energy from the environment that breaks Cooper-pairs (creating \textit{quasiparticles}), the resultant change in kinetic inductance modulates the complex transmission $S_{21}$ of a microwave frequency tone. This modulation can be interpreted as a shift of the resonant frequency $\nu_r$ and the internal quality factor $Q_i$ of the resonator. Aided by the high quality factors achievable by superconducting devices ($Q_i>10^6$ \cite{hammer2008}), hundreds to thousands of narrow linewidth MKIDs of $\nu_r=$~\bigo{GHz} can be chained on a single RF feedline with \bigo{MHz} spacing \cite{shu2018increased}, providing for straightforward room temperature readout and operation. MKIDs have been used for numerous astrophysical applications ranging from sub-millimeter wavelength telescopes to optical exoplanet observatories \cite{{ulbricht2015highly,adam2018nika2,golwala2012,meeker2018}}. 

The past decade has also seen the development of lumped-element\footnote{Rather than distributed inductive and capacitive elements.} MKIDs \cite{doyle2008lumped} for use as Kinetic Inductance Phonon-Mediated detectors (KIPMs). In this concept, incident particles on a crystalline substrate (like silicon) create propagating lattice vibrations (\textit{athermal phonons}) that couple to the MKID sensors patterned on the substrate's surfaces. With large gram to kilogram mass targets, these devices can be used as particle detectors for rare-event searches. For example, this phonon-mediated scheme has been demonstrated for x-ray detection \cite{moore2012}, for neutrinoless double-beta decay searches \cite{cardani2021}, and as future dark matter detectors \cite{ramanathan2022, wen2022}.

The effectiveness of KIPM detectors for particle physics applications hinges on their ability to reconstruct energies $E$ with the smallest possible baseline energy resolution $\sigma_{E}$. From the derived expressions in Refs. \cite{zmuidzinas2012,moore2012} we expect the energy resolution to scale as,
\begin{equation}
    \sigma_{E} \propto \sqrt{T_{\rm sys}}\\
    \label{eq:resolution} 
\end{equation}
in the case of a white noise \textit{amplifier-limited} readout with overall system noise temperature $T_{\rm sys}$. Two level system (TLS) noise, commonly arising from surface oxides, is another important source of noise in MKIDs and is expected to be sub-dominant in designs with large feature sizes of the capacitor \cite{noroozian2009}.  Recent KIPM detector results have inferred a 5~eV energy resolution for an amplifier-limited detector and a 20~eV energy resolution for a TLS-limited detector \cite{wen2022}. At cryogenic millikelvin temperatures where temperature effects are non-linear \cite{nyquist1928thermal}, noise at a frequency $\nu$ and temperature $T$ can be better expressed as \textit{noise quanta} (i.e. photons) \cite{caves1982} via 
\begin{equation} \label{eq:N}
    N = \frac{1}{2}{\rm coth}\left(\frac{h\nu}{2 k_B T}\right),
\end{equation}
per unit bandwidth. This quantity captures the quantum mechanical behavior of noise that can manifest at very low temperatures where $T\leq h\nu/ k_B$. In this Letter, we demonstrate a strategy to minimize the system noise, $N_{\rm sys}$, across a wide 70~MHz band, and compare it to the Standard Quantum Limit (SQL) of $N=1$\footnote{A combination of 1/2 photon zero point vacuum fluctuations and 1/2 photon amplifier added noise. The SQL technically refers to the amplifier added photon contribution, but here we consider the effective noise bound for an actual experiment.}, the quantum mechanical lower bound of any experiment using signal amplification\footnote{Though techniques like ``squeezing" may help surmount this bound \cite{caves1981quantum, esposito2022observation}} \cite{caves1982, aumentado2020superconducting}. 

With their large gain ($G>30$~dB) and complementary wide bandwidth, high electron mobility transistor (HEMT) amplifiers are the dominant amplification technology used by present MKID experiments for simultaneous readout of all sensors. However, with nominal noise temperatures of $\mathcal{O}$(K) at C-band frequencies (4\textendash8~GHz), these devices add $\mathcal{O}(10)$ quanta of noise to any measurement. As Eq. \ref{eq:resolution} would dictate then, a theoretical improvement of $>4 \times$ the resolution would be achievable through quantum limited operation.  

Kinetic Inductance Traveling Wave Parametric Amplifiers (KI-TWPAs) are a class of quantum limited amplifiers which also exploit principles of kinetic inductance, but for exactly such low-noise amplification \cite{ho2012}. They have been demonstrated to reach the SQL limit of added amplifier noise \cite{{klimovich2022, klimovich2023}} and have previously been integrated with optical MKIDs \cite{zobrist2019}. In this paper we demonstrate the noise improvement achievable in a KIPM detector that is coupled to a KI-TWPA that offers broadband gain and low noise performance for future rare-event search experiments.

\section{Setup and Measurement} \label{sec:setup}

\begin{figure*}[!ht]
	\includegraphics[width=0.99\textwidth]{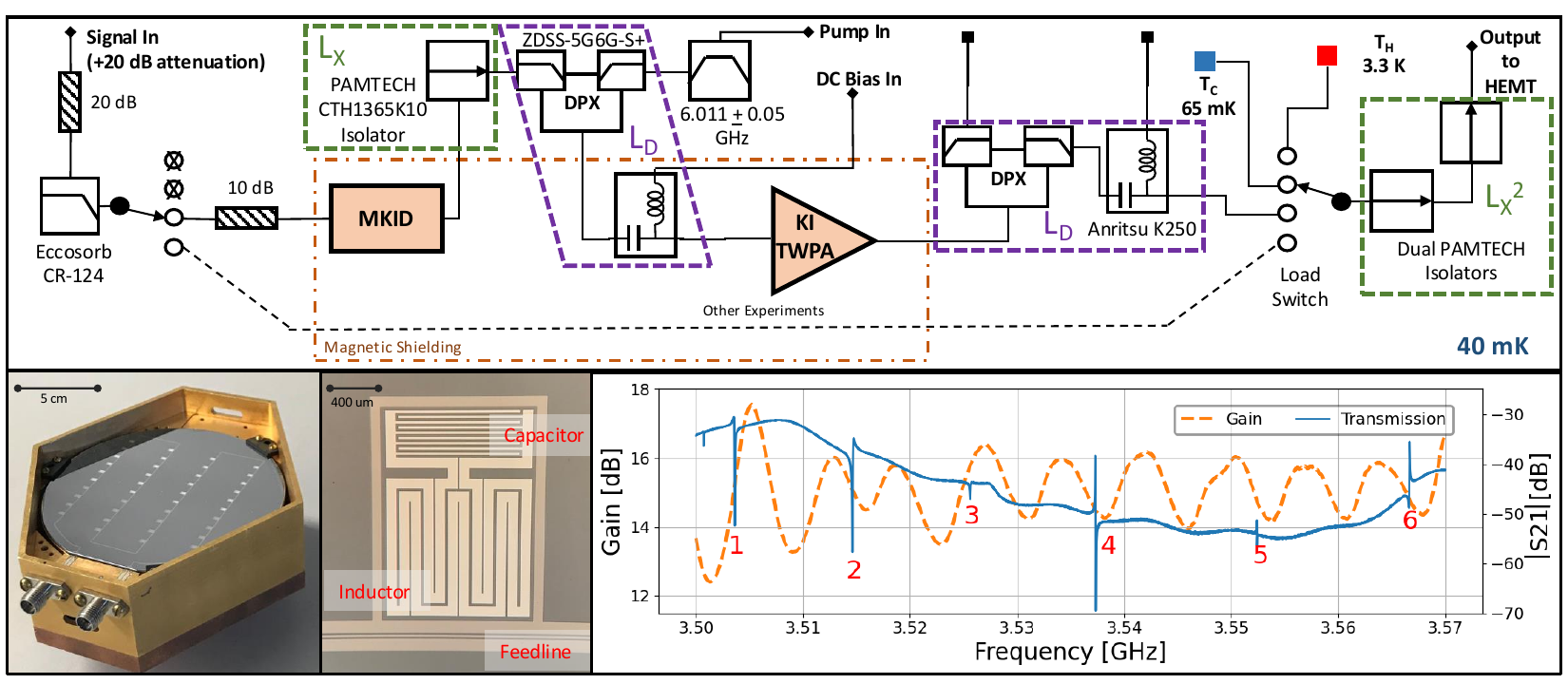}
    \caption{\label{fig:schematic} \textit{Top:} Schematic of cold stage of experiment, with labeled components. Thick dashed lines indicate groupings labeled by their losses $L_D$, $L_X$, and $L_X^2$ of passive components referred to in Sec. \ref{sec:calibration}. Eccosorb CR-124 is a microwave-absorbing epoxy, used as a GHz-scale low-pass filter to attenuate stray radiation at the device input. \textit{Bottom Left: } Picture of 40 MKID array used in experiment showing the gold plated copper device holder, with cirlex clamped silicon substrate, along with a micrograph of a single MKID, outlining the constituent circuit features. \textit{Bottom Right:} Transmission $S_{21}$ of the MKID array, showing the numbered Al resonators used in the analysis (leftmost resonance feature is a Nb resonator). Overlaid is the smoothed gain, $G_{\rm pa}$, of KI-TWPA in detector band, as measured between KI-TWPA on and off conditions.}
\end{figure*}

The cold stage components of the experiment are laid out in the schematic of Fig.~\ref{fig:schematic} Top. The MKID array used in this experiment was designed and fabricated based on parameters and principles found in Ref. \cite{chang2023}. It has 40 sensors, 16 made of 30~nm thick Aluminum and 24 made of 30~nm Niobium, coupled to a 300~nm thick Nb feedline and deposited on a 3~inch wafer as seen in Fig.~\ref{fig:schematic} Bottom Left. The sensors span $\nu_r$ from approx. 3.0~GHz to 3.6~GHz and have total quality factors $Q_r$ ($=1/Q_i+1/Q_c$, where $Q_c$ is the coupling quality factor) between 10$^3$\textendash10$^5$, with the variation in $Q_r$ likely due to the large differences in $S_{21}$ transmission across the operating band \cite{ramanathan2022}, as seen in Fig.~\ref{fig:schematic} Bottom Right, along with other effects such as Fano interference \cite{Rieger2023}. The variation in $S_{21}$ is perhaps due to spurious ``box-modes'' arising from electromagnetic coupling between the device and the enclosure, as described in Appendix \ref{app:KIPMnonideal}. For this exercise we operated the KIPM detector in a dilution refrigerator at $T_0 = 40\pm1$~mK and collected data from 6 Al MKIDs in the 3.5\textendash3.6 GHz range, using individual readout tones generated by a room-temperature signal generator with cold stage attenuation and filtering to reduce input noise. Aluminum films are of interest as phonon sensors due to their phonon collection efficiency, their low gap energy ($\Delta \sim 200$~$\mu$eV), and their long quasiparticle lifetimes \cite{moore2012, kaplan1976}. These specific MKID sensors were chosen because they had the highest off-resonance feedline transmission amongst all candidates, ensuring maximal readout power delivered to each sensor.. The array was connected to a NbTiN KI-TWPA operated in 3-wave mixing mode (operational details found in Ref. \cite{{klimovich2022, klimovich2023}}), with associated passive diplexers and bias tees for introducing and dumping the 6.11~GHz pump tone and DC bias current $I_{\rm DC}$ required for the KI-TWPA's operation. In addition, the setup used 3 identical isolators to reduce harmful reflections. The output was fed to a cold switch, with other throws linked to a ``hot" ($3.38\pm0.05$~K) and ``cold" ($65\pm1$~mK) load (acting as Johnson noise sources realized via 50~$\Omega$ coaxial terminations thermally anchored to fridge temperature stages) for calibration purposes. The output was further amplified by a HEMT (LNF-LNC03.14A with a vendor specified gain of 39~dB and noise temperature of 3.6~K at 3.5~GHz) and a 27~dB room temperature amplifier. Signal processing was done using either a homodyne setup (IQ mixer fed to a PCI-E DAQ card) or direct connection to a spectrum analyzer (SA) / vector network analyzer (VNA) for calibration.

Three separate measurements were performed, allowing determination of the KI-TWPA gain, the system noise, and individual resonator noise. The KI-TWPA gain was measured using the VNA and measuring the difference in transmission with the DC bias and pump tone turned on/off. The measured $\sim$15~dB gain profile ($G_{\rm pa}$) of the KI-TWPA is shown in Fig.~\ref{fig:schematic} Bottom Right. The relatively low gain is attributed to a smaller than expected critical current parameter for the specific device, not allowing it to be driven with a larger pump \& bias. The observed ripples, of $\mathcal{O}$(MHz) periodicity, are likely sourced from an impedance mismatch between the parametric amplifier and other circuitry, as described in Ref. \cite{{klimovich2022, klimovich2023}}, and we discuss some features of these ripples in Appendix \ref{app:KITWPAnonideal}. 

The calibration of the added noise and system noise temperature was done using the aforementioned cold-switch setup with the output signal processed by a real-time spectrum analyzer. 4 power data sets were collected in this configuration: a 300~K terminated input with the KI-TWPA on ($P_{\rm on}$) and off ($P_{\rm off}$), along with the hot load ($P_{\rm H}$) and cold load ($P_{\rm C}$) measurements.

The resonator based measurements first used the VNA to perform a $\nu_r \pm 5$~MHz characterization sweep of each resonator's complex $S_{21}$ resonance circle. Next, we used the DAQ setup to record $S_{21}$ timestreams from each of the 6 resonators, using signal generator created tones. For each resonator we recorded 10~s long timestreams at their respective $\nu_r$, along with 0.5~s timestreams at 12 logarithmically spaced frequency intervals around $\pm5\nu_r/Q_r$ including two far off resonance datasets at $\nu_r\pm20\nu_r/Q_r$. These measurements were repeated at three powers $P_g=[-80,-68,-56]$~dBm (at the input port of the KIPM array, after cold-stage attenuation) as well as for the KI-TWPA on and off conditions. All outputs were recorded at a sampling rate of 700~kS/s per quadrature and further digitally low-pass filtered and downsampled by a factor of 10$\times$ to 70~kS/s per quadrature. This is sufficient to capture physically relevant pulse information as the resonator bandwidth is expected to be $\nu_r/(2Q_r)$ $\sim$ 20 kHz, based on future KIPM designs targeting a $Q_r$ limiting $Q_c \sim\,10^{5}$.

\section{System noise calibration} \label{sec:calibration}

With the gain measurement in hand we computed the resolution improvement in two independent ways, either using the spectrum analyzer or the individual resonator timestream datasets described above. In this section we detail the spectrum analyzer method, where we performed a two-stage analysis to compute the system noises $N_{\rm sys}^{\rm on}$ and $N_{\rm sys}^{\rm off}$ at the output of the MKID (before the first isolator), for the KI-TWPA on and off conditions respectively. We first computed the post-switch isolator and HEMT noise contribution using a 'Y-factor' analysis, and then folded in the pre-switch isolators, diplexers, and KI-TWPA. Calculations were performed referencing the switch plane, since the existence of measured thermal loads at that location allowed for noise comparisons, before being flowed back to the MKID output plane. We refer to the groupings of passive elements from Fig.~\ref{fig:schematic} Top (see thick dashed purple and green lines) and have generally follow the prescriptions of Refs. \cite{{zobrist2019,klimovich2022, klimovich2023}} for specific calculations. Throughout, gains $G>1$ and losses $L>1$ are specified as dimensionless linear power ratios for components that amplify or attenuate their inputs respectively. For example, $L = 10^{|L_{\rm dB}|/10} > 1$ for a component with insertion loss $L_{\rm dB}<0$~dB. 

The Y-factor method determines the noise temperature of an amplifier chain without requiring knowledge of its absolute gain or bandwidth. Generically, the output power of a system viewing an input temperature $T$ with added noise $T_{\rm noise}$ is $P_{\rm out}=Gk_BB(T+T_{\rm noise})$, for system gain $G$ and bandwidth $B$. By measuring the output power at the analyzer one can compute the power ratio $Y=P_H/P_C$, as a function of frequency, when the input is switched between a hot load at known temperature $T_H$ and a cold load at known temperature $T_C$ respectively. This ratio cancels out $G$ and $B$, allowing $T_{\rm noise}$ to be determined and expressed solely in terms of $T_H$, $T_C$, and $Y$. We used this prescription to write the effective noise temperature of the post-switch isolators and HEMT package as
\begin{gather} \label{eq:HEMT}
    T^{\rm Isolators+HEMT} \approx \frac{T_H' - Y T_C'}{(Y-1)}= 10.4~\rm{K}\:(\pm0.7~\rm{K}),
\end{gather}
evaluated in the frequency band spanning the resonators. Here we have identified $T_{\rm noise}\equiv T^{\rm Isolators+HEMT}$ in the Y-factor calculation, as the HEMT has nearly 40~dB worth of gain (i.e. 10000$\times$ amplification) and thus dominates everything downstream of it\footnote{Downstream components will contribute to the noise temperature as $T_{\rm component}/G_{\rm HEMT}$}. Note in this calculation we have used primed ``effective" temperatures for the hot and cold loads, where $T' = (h\nu/2k_B)\coth(h\nu/2k_BT)$, an application of Eq. \ref{eq:N} to convert temperature to a noise quanta basis and then back again to an effective linearized temperature. The resulting $T^{\rm Isolators+HEMT}$ is greatly elevated from a HEMT only datasheet value of nominally 3.6~K, suggesting that the isolators have significantly worsened the noise.

Next, we include the contribution of the pre-switch components including the parametric amplifier. As previously stated, since the switch plane has a defined physical temperature noise source, we can compare the measured power at the analyzer for the KI-TWPA on and off ($P_{\rm on}$ and $P_{\rm off}$ respectively, achieved by turning on/off the KI-TWPA pump tone) to the known points of ($T_C'$,$P_C$) and ($T_H'$,$P_H$) to determine effective $T_{\rm on/off}$ quantities. The known temperature and power points helped define the linear relationship $P=m\cdot T + y_0$, with $m$ slope and $y_0$ intercept. Thus, the noise temperature of the component chain before the switch was expressed at the switch plane as $T_{\rm on/off}=(P_{\rm on/off}-y_0)/m$. 

Finally then, the overall system noise at the output of the MKID, expressed in noise quanta, was computed for KI-TWPA on and off as
\begin{align} \label{eq:finalnoise}
    N_{\rm sys}^{\rm on/off} \approx \bigg( T^{\rm Isolators+HEMT}+\frac{P_{\rm on/off} - y_0}{m} \bigg) \bigg(\frac{L_D^2 L_X}{G^{\rm on/off}_{\rm pa}} \bigg) \bigg( \frac{k_B}{h\nu} \bigg)
\end{align}
where the first bracketed term is the sum of the two computed noise temperatures. The second bracketed term encapsulates the gains and losses of the components upstream of the switch (2 diplexers, 1 isolator, and the KI-TWPA) and, as per Friis noise convention, are factors used to move the reference plane from the switch back to the MKID output. The third bracketed term converts the result from noise temperature to noise quanta.

\begin{figure}[!ht]
    \includegraphics[width=0.99\textwidth]{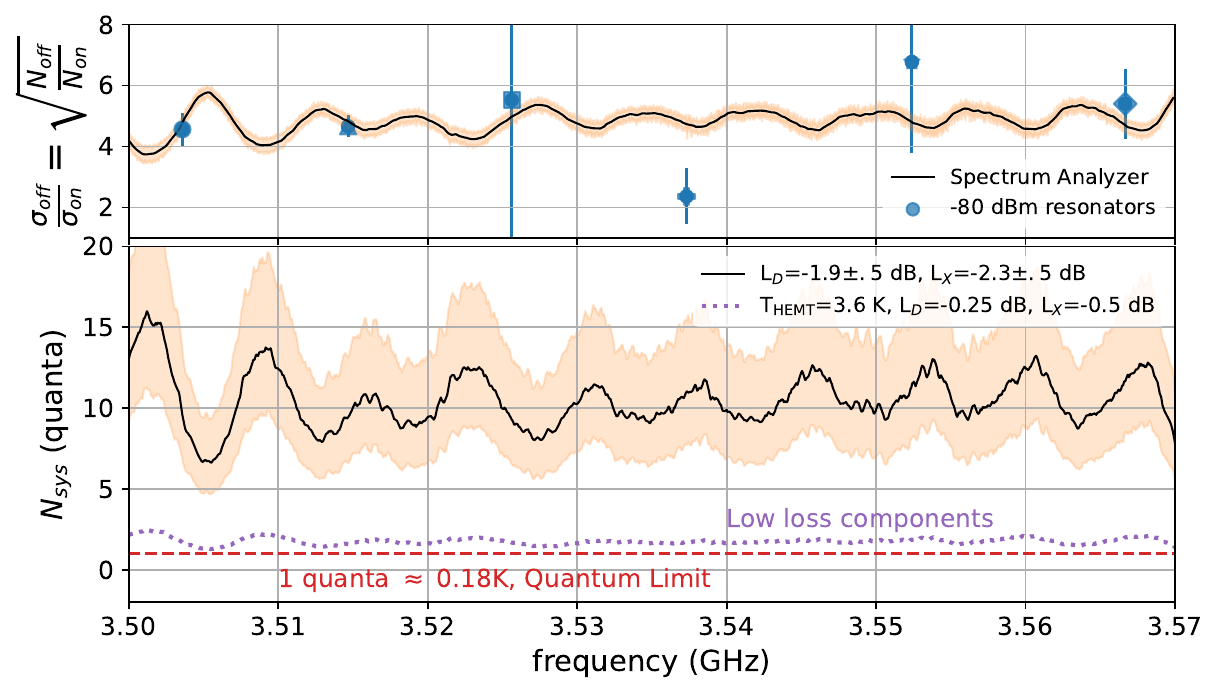}
    \caption{\label{fig:improvement} \textit{Top:} Resolution improvement as observed from both the spectrum analyzer and individual lowest power resonator data, with $\pm1\sigma$ power measurement uncertainty shown by the shaded band. \textit{Bottom:} Total system noise quanta with the KI-TWPA on, as referred to the input of the first isolator. The shaded region reflects systematics on the passive component attenuations sandwiching the KI-TWPA (see Fig.~\ref{fig:schematic}).  Using datasheet and prior measurement values results in the solid (black) line, indicating about 10~quanta of added noise with a minima around 7 quanta. The dotted purple line shows near SQL performance is achievable with better cold circuitry.}
\end{figure}

Note that this calculation is the system noise as it pertains to the entire experiment \textemdash\ exactly the configuration when performing a rare-event search \textemdash\ and not just the added contribution from an SQL amplifier. We plot the expected resolution improvement as the root ratio of the on/off conditions of Eq. \ref{eq:finalnoise}, as seen by the solid line in Fig.~\ref{fig:improvement} Top, indicating an approximately 5$\times$ improvement in energy resolution. The 1$\sigma$ uncertainty band expressed there arises from variations within the power measurements affecting $Y$ and $P_{\rm on/off}$. The resonator points will be discussed in Sec. \ref{sec:resonator}.

From datasheets and prior measurements we estimated the diplexer + bias-tee attenuation as $L_D=-1.9$~dB and the isolator $L_X=-2.2$~dB \cite{{diplexermanual,btmanual,isolatortest}}, leading to the system noise curve in Fig.~\ref{fig:improvement} Bottom (along with systematic uncertainties), averaging $N_{\rm sys}\approx10$.\footnote{These passive components are not precisely matched to our device frequency band, leading to larger than intended insertion losses.} The latest generation of commercial HEMT amplifiers can achieve noise temperatures of $\sim$1.5~K (e.g. the LNF-LNC4.8G) which also suggests $N_{\rm sys} \approx$ 10, casting doubt on the usefulness of a parametric amplifier for a rare-event search experiment. We resolve this tension by showing the effect of lossy passive components on the result. By selecting for optimal commercially available/achievable $L_D=-0.25$~dB and $L_X=-0.5$~dB and \textit{fixing} $T_{\rm HEMT}=3.6$~K (our datasheet value), we can flow these changes through Eqs. \ref{eq:HEMT} and \ref{eq:finalnoise}, leading to the purple dotted line close to the SQL in Fig. \ref{fig:improvement} bottom. This indicates that the dominant loss terms arguably arise from our passive components. Thus, further improvements should come both from refined operation of the KI-TWPA with higher gain ($G_{\rm pa}\gtrapprox30$~dB \cite{klimovich2023}) and in carefully selecting for far less lossy passive components. We note however that an unmodeled additional noise source along the chain, perhaps coming from residual thermal radiation through a switch port, would similarly degrade the inferred system noise floor and cannot be ruled out. However, Howe et al. \cite{howe2025compact} have recently built a traveling wave amplifier with on-chip RF components (the ``ORCK" platform) and showed a median system noise of just 3.4 quanta (which includes the effect of multiple passive components), indicating some level of improvement is immediately feasible. Finally, in a real rare-event search where the KIPM is to be calibrated, one can remove both switches in Fig.~\ref{fig:schematic} at the expense of precise amplifier calibration.  

\section{Resonator noise calibration}\label{sec:resonator}

In this section, we detail the operation of the KIPM resonators with the KI-TWPA on and off, and demonstrate the noise improvement achievable for an actual experiment. Details of the readout scheme are summarized in Sec. \ref{sec:setup}.

The complex VNA transmission data, $S_{21}$, of each resonator (an example of the modulus $|S_{21}|$ across the entire band is shown in Fig.~\ref{fig:schematic} bottom right) was fit to the following form from Ref. \cite{khalil2012}:
\begin{equation}
    S_{21}(\nu) = a \cdot e^{2 \pi i \nu \tau}\bigg ( 1 - \frac{Q_r|\hat{Q}_c|^{-1}e^{i\phi}}{1 + 2 i Q_r x} \bigg) 
    \label{eq:s21}
\end{equation}
for complex coupling quality factor $\hat{Q}_c$\footnote{$\hat{Q}_c$ is now complex to account for impedance mismatches along the feedline.} parameterized in terms of its inverse magnitude and phase $\phi$, complex off-resonance transmission $a$, cable delay $\tau$, and resonant frequency detuning $x=(\nu - \nu_r)/\nu_r$. The varying linewidths and line-shapes of the resonators is believed to arise from impedance mismatches either caused by spurious EM modes from wiring, feedline geometry, or the enclosure (see Appendix \ref{app:KIPMnonideal}).

\begin{figure*}[!ht]
    \centering
    \includegraphics[width=.99\textwidth,trim={1.25cm 0 1.25cm 0}]{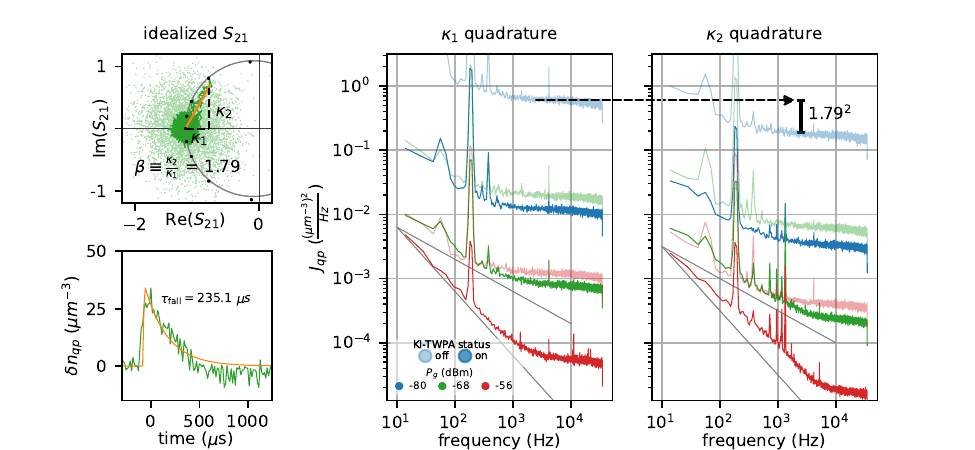}
    \caption{\label{fig:IQ} \textit{Top Left:} The idealized $S_{21}$ timestream of MKID 5 at -68~dBm, showing KI-TWPA on (off) data points in solid (faded), overlain on the unit normalized resonance circle (grey) with parameters derived from the fits to the VNA scan. The black points are the means of the IQ timestreams with the KI-TWPA on, measured at logarithmically spaced frequencies around $\nu_r$ and converted into idealized $S_{21}$. The green deviation is an observed pulse, and the orange line represents the expected pulse direction as given by Mattis-Bardeen theory. \textit{Bottom Left: } Extracted change in quasiparticle density from the same pulse, with overlain single exponential decay fit. \textit{Center, Right:} $J_{\rm qp}$ for MKID 5 at various powers and with the KI-TWPA on and off, to illustrate the improvement in the noise with the KI-TWPA. $f^{-1}$ and $f^{-0.5}$ have been drawn to guide the eye.} 
\end{figure*}

After the raw VNA data is fit to \ref{eq:s21}, the extracted parameters can be used to transform the data into an idealized $S_{21}$:
\begin{equation}
S_{21,\text{ ideal}}(\nu) = 1 - \frac{Q_r}{Q_c}\frac{1}{1+2iQ_rx},
\end{equation}
where $Q_c^{-1}=|\hat{Q}_c|^{-1}\cos{\phi}$. Note the inverted nature of the cosine projection. In words, the idealized transmission $S_{21,\text{ ideal}}$ divides out the cable delay $\tau$ and off-resonance transmission $a$ as well as projects the resonance circle onto the real-axis, eliminating $\phi$. The noise timestreams, both on- and off-resonance, are also corrected in this way. The idealized versions of the VNA data and overlain timestream data from MKID \#5 are shown in Fig.~\ref{fig:IQ} top left\footnote{We expect $Q_r<Q_c$ and thus $\operatorname{Re}[S_{21,\text{ ideal}}]>0$. That is not observed in Fig.~\ref{fig:IQ} top left. In the raw data, the circle is significantly rotated, so the simple interpretation of impedance mismatches as per Ref. \cite{khalil2012} may not be complete. See Appendix \ref{app:KIPMnonideal} for further discussion.}. The improvement in readout noise can be seen directly as a decrease in the scatter of the noise points. 

% OW commented out and re-phrased above
% The I and Q quadratures of the on resonance raw timestream $S_{21}$ data ($=I+iQ$) acquired using the DAQ were converted into an idealized form by using the VNA fit parameters and the logarithmically spaced IQ timestreams. Specifically, the raw data are corrected by dividing out the cable delay $\tau$ and off-resonance transmission $a$ terms in Eq. \ref{eq:s21}, followed by a projection onto the real-axis to eliminate $\phi$ \textcolor{red}{what was done}. An example of the on-resonance idealized $S_{21}$ data is displayed with the KI-TWPA both on and off in Fig.~\ref{fig:IQ}, overlaid on the unit normalized resonance circle. The improvement in readout noise can be seen directly as a decrease in the scatter of the noise points.

Next, deviations about the mean idealized $S_{21}$ were computed as a function of time $t$ and cast as $\delta S_{21}(t)$. The timestreams were converted into a \textit{quasiparticle basis} ($\kappa_1$, $\kappa_2$) using the relation,
\begin{equation}
    \delta S_{21}(t) =\alpha \frac{Q_r^2}{Q_c}(\kappa_1 + i\kappa_2)\delta n_{\mathrm{qp}}(t),
    \label{eq:k1k2}
\end{equation}
where $\alpha$ is the kinetic inductance fraction of the film and $\kappa_{1,2}(T_0,\nu_r,\Delta)$ are derived from \textit{Mattis-Bardeen} theory \cite{mattis1958theory} evaluated at base temperature $T_0$=40~mK, and represent the film conductivity response to changing quasiparticle density $n_{\rm qp}$. For small perturbations in the density, these orthogonal deviations can be used to identify $\delta n_{\rm qp}$. 

To obtain $\alpha$ and $\Delta$, the temperature $T$ was swept up to 350~mK, and $\nu_r(T)$ and $Q_r(T)$ were measured for each resonator. In general, either quadrature may be used to extract $\alpha$ and $\Delta$, but only $\nu_r(T)$ was used in this analysis. $Q_r(T)$ showed evidence of non-thermal sources of dissipation, such as idealized resonance circles that encircle 0 (an example of which is in Fig.~\ref{fig:IQ} Top Left), and consequent deviations from Mattis-Bardeen expectations, which prevented reliable extraction of $\alpha$ and $\Delta$ in the $\kappa_1$ quadrature. Possible sources of non-thermal sources of dissipation are gap broadening \cite{zmuidzinas2012, Mitrovic_2008} and vortex loss \cite{Stan_2004}.

\begin{figure*}[!ht]
    \centering
    \includegraphics[width=.99\textwidth,trim={0.5cm 0 0.5cm 0}]{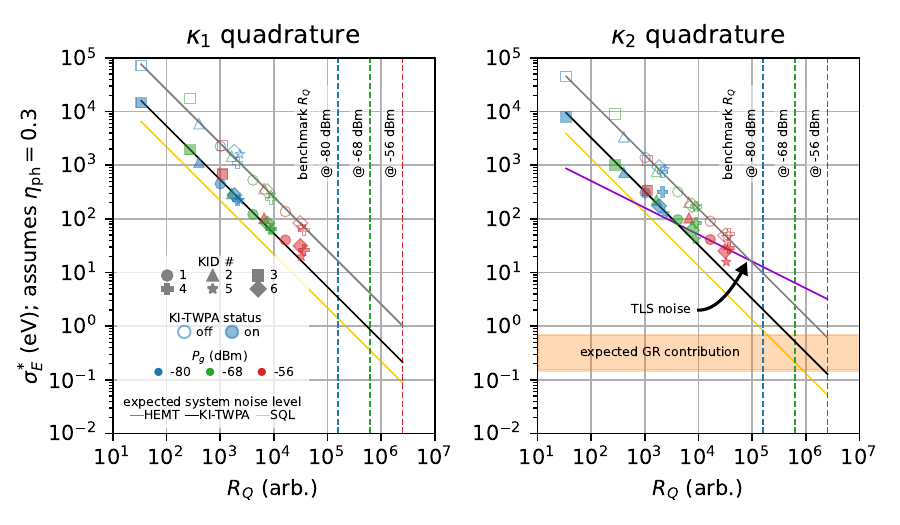}
    \vspace{-0.5cm}
    \caption{\label{fig:rqplot} Inferred energy resolution across orthogonal quadratures, resonator number, readout power, and operating configuration, plotted against the RF responsivity $R_Q$ defined in Eq. \ref{eq:rq}, along with expected system noise levels for different scenarios. The vertical lines at different powers reflect benchmark $R_Q$ quantities as discussed in text. In the $\kappa_2$ quadrature, the effect of TLS noise, expected to go as $P_g^{-1/4}$, is drawn suggestively in purple. Expected generation-recombination noise resolution limits (applicable to both quadratures, but shown only in $\kappa_2$) are illustrated by the shaded orange band.}
\end{figure*}

The relative responsivities of the $\kappa_{1}$ and $\kappa_2$ quadratures can be expressed quantitatively as $\beta \equiv \kappa_2(T_0,\nu_r,\Delta)/\kappa_1(T_0,\nu_r,\Delta)$. In principle, $\beta$ can be predicted by Mattis-Bardeen theory; see Ref. \cite{siegel2016multiwavelength} for theory curves. In practice, it can be measured using the temperature sweeps as described previously or using the pulse shape of the $S_{21}$ trajectory when incident energy is received by a particular MKID. In Fig.~\ref{fig:IQ} Top Left, the pulse shape of an \textit{in situ} observed particle event displays good agreement with the theoretical expectation for $\beta$, shown in orange. This agreement was not seen for all of the observed particle events in the various timestreams, of which there were only a handful, likely due to non-thermal sources of dissipation as mentioned previously. Further study of $\beta$ will require a dedicated energy source, such as a radioactive material or an LED, so as to improve statistics. For the purposes of this article, we assume the theoretical Mattis-Bardeen response to define the relative responsivity of the two quadratures.

An example of a timestream converted into quasiparticle units is shown for the $\kappa_2$ direction in Fig \ref{fig:IQ} Bottom Left and is the same pulse that is shown in Fig \ref{fig:IQ} Top Left. A single exponential decay, $\delta n_{\rm qp} \propto \rm{exp}(-t/\tau_{\rm fall})$, was fit to this observed pulse and the fall time constant is displayed. The fit was then used to construct a signal template $\tilde{s}$, in frequency space, which in conjunction with a noise power spectral density $J_{\rm qp}$ was used to compute the optimally filtered resolution on quasiparticle density 
\begin{equation}
    \sigma_{\rm qp}^2 = \left[\int_{-\infty}^{\infty}df \frac{|\tilde{s}(f)|^2}{J_{\rm qp}(f)} \right]^{-1}
    \label{eq:optimal}
\end{equation} 
as per the standard prescription found in Refs. \cite{wen2022, golwala2000}. Briefly, the optimal filter formalism is a low-pass filter that weights the signal so as to both cut away noise contributions from higher frequencies that do not contribute to the shape of the signal, and down-weights any low frequency rise in the noise.

$J_{\rm qp}$ was computed for both $\kappa_{1,2}$ timestreams, and examples for MKID 5 are plotted in Fig.~\ref{fig:IQ} Center and Right. The effect of the theoretically specified $\beta$ on the relative noise level between quadratures is illustrated with the dashed arrow. The impact of the KI-TWPA on the readout noise is seen across all readout powers at the highest frequencies: around a decade-and-a-half of improvement, matching the 5$\times$ system noise improvement from the spectrum analyzer measurement. At lower frequencies, both quadratures demonstrate significant noise coloring, progressively increasing with  readout power. In the case of $\kappa_2$, the resonator's intrinsic TLS noise dominates and exhibits a signature frequency dependence $f^{-1\leq y\leq-0.5}$ below the roll-off frequency $\nu_r/(2Q_r)$ \cite{gao2008physics}. The exponent $y$ is in general material and fabrication process dependent and given by the oscillatory behavior of the two-level system. TLS noise of this form has been previously noted as limiting the noise performance of similarly designed MKIDs \cite{wen2022}. The $\kappa_1$ quadrature is not expected to exhibit TLS \cite{gao2008physics}, and the noise follows a $f^{-1}$ prescription at the highest readout power. One hypothesis for this noise is that it is sourced by long timescale transmission fluctuations. If correct, this noise may be removable in the future by the concurrent measurement of an off-resonance tone and subsequent subtraction of any correlated component between the on-resonance tone and the off-resonance tone. Similar techniques have been demonstrated for noise reduction in prior MKID measurements \cite{ramanathan2022, wen2022}.

%$\tilde{s}$ and $J_{\rm qp}$ were used to compute the optimally filtered quasiparticle resolution $\sigma_{\rm qp}$
With $\tilde{s}$ and $J_{\rm qp}$ in hand, we computed the optimally filtered quasiparticle density resolution $\sigma_{\rm qp}$ as per Eq. \ref{eq:optimal} and the results are shown in Fig.~\ref{fig:rqplot}. In order to project the ultimate energy threshold reach of a KIPM detector with KI-TWPA amplification, the results are converted to an inferred energy resolution $\sigma_E^*$ of a hypothetical single resonator KIPM detector\footnote{In a multi-resonator KIPM detector, the energy resolution scales with $\sqrt{\text{\# of resonators}}$.}, assuming the measured $\sigma_{\rm qp}$ for each MKID in the present array. We converted to $\sigma_E^*$ with
\begin{equation}
\sigma_E^* = \frac{ V_{\rm sc}\Delta}{\eta_{\rm ph}}\sigma_{\rm qp}\,,
\label{eq:sigE}
\end{equation}
where $V_{\rm sc}=3\times10^4$~$\mu$m$^3$ is the superconducting absorber volume and $\eta_{\rm ph}$ is a phonon absorption to quasiparticle creation efficiency. We expect $\eta_{\rm ph}=0.3$ is achievable, from measurements in other phonon-mediated experimental designs \cite{ren2021design,pyle2009surface}.

The results of the inferred resolution calculation are shown in Fig.~\ref{fig:rqplot} for both the $\kappa_{1,2}$ directions, plotted against an \textit{RF responsivity} quantity $R_Q$. This variable, first defined in Ref. \cite{ramanathan2022}, takes the form:
\begin{equation}
    R_Q \equiv |V_r|\frac{Q_r^2}{|\hat{Q}_c|}\,,
    \label{eq:rq}
\end{equation}
where $|V_r|$ is the voltage amplitude of a nearby off-resonant tone \textit{at the resonator}. Given the complexity and systematic uncertainties involved in extracting $V_{r}$ in a real experiment, a relative measure of $R_Q$ can be obtained by using $V_{\rm ADC}$ instead, which is measured at the DAQ in arbitrary ADC units. Conveniently, the diameter of a resonance circle in raw IQ space is $D=|V_{\rm ADC}|Q_r/|\hat{Q}_c|$, which allowed for this relative $R_Q$ to be computed as $DQ_r$. We note that $|V_{\rm ADC}| \propto |a|\sqrt{P_\text{DAC}}$: $a$~is the off-resonance transmission, introduced in Eq.~\ref{eq:s21}, and $P_\text{DAC}$ is the power at the DAC, which is also proportional to $P_g$, the generator or readout power at the resonator. In other words, as the readout power increases, $|V_\text{ADC}|$, $|V_r|$, and ultimately $R_Q$ all increase, leading to an improved energy resolution. Additionally, $R_Q\propto dS_{21}/dx|_{x=0}$ as computed from Eq. \ref{eq:s21}, emphasizing the role of $R_Q$ as the RF responsivity.

Fig.~\ref{fig:rqplot} shows that $R_Q$ dominates the energy performance of the investigated resonators due to the wide range of transmission and quality factors that were measured across the resonators. Two orders of magnitude may be gained in the energy resolution by rectifying $R_Q$ of the resonator design to hit our displayed benchmark for all sensors. The design goal is a device transmission of unity and $Q_r\sim Q_c\sim 10^5$ so that $\mathcal{O}(\mu\rm{s})$ rise time information can be maintained, with ongoing efforts to improve $R_Q$ using RF engineering techniques \cite{kim2009wire}. Higher readout powers are also beneficial but have the caveat of affecting detector linearity \cite{siegel2016multiwavelength}.

Fig.~\ref{fig:rqplot} also shows the inferred energy resolution improvement from KI-TWPA operation as the level shift between the KI-TWPA on and off points. For the KI-TWPA off points, $R_Q$ as measured at the DAQ was multiplied by $G_{\rm pa}$ to put it on the same footing as the KI-TWPA on scenario. This step ensured the $R_Q\propto|V_r|$ scaling. The resolution improvement is also recorded as the scatter points in Fig.~\ref{fig:improvement} Top, showing good alignment with the spectrum analyzer measurements. At low readout powers, the conversion to the idealized $S_{21}$ has an uncertainty that is dominated by the size estimation of the resonance circle in the IQ plane, and it is this uncertainty that is propagated into the error bars seen in Fig.~\ref{fig:improvement}. The expected system noise levels are illustrated via the solid lines that pass through the points: the HEMT-dominated system noise level is determined from a fit to the points with the KI-TWPA off, and the expected system noise levels of the KI-TWPA-dominated and the SQL cases come from the spectrum analyzer analysis. 

At the lowest readout powers and for resonators of low $R_Q$, the full range of expected improvement is observed. At higher readout powers, two stories emerge: 1) in the $\kappa_1$ quadrature, the aforementioned $f^{-1}$ noise begins to contribute more; 2) in the $\kappa_2$ quadrature, TLS noise begins to dominate the noise performance. While no complete microphysical description of TLS noise exists, semi-empirical work by Ref. \cite{gao2008physics} has demonstrated that TLS noise spectral density evolves as $J_{\rm TLS} \propto P_g^{-1/2}$ (again, the readout power). Feeding this limit into Eq. \ref{eq:optimal}, yields $\sigma_{\rm qp}\propto P_g^{-1/4}$, and the points at higher readout powers in Fig.~\ref{fig:rqplot} Right follow this relation. It is believed that this particular MKID, \#4, exhibits elevated TLS noise due to the presence of multiple oxide-forming films during the fabrication process. Crucially then, the resolution improvement due to parametric amplification is dampened by these effects, visible for that specific device in Fig. \ref{fig:improvement}, that will have to be solved to realize a quantum-limited sensor at higher readout powers. Furthermore, the ultimate achievable resolution performance of SQL amplification is still a factor of $\sqrt{\langle{N_{\rm sys}}\rangle} \sim 3\times$ below current performance, likely limited by the prior discussed passive component systematics and KI-TWPA gain. 

One unobserved noise source in this analysis that will be important going forward is generation-recombination (GR) noise of the quiescent quasiparticle population (density $n_0$). The observed lifetime from Fig.~\ref{fig:IQ} Bottom Left implies $n_0 \approx 475\;\mu\mathrm{m}^{-3}$ \cite{kaplan1976}. Using the current design $V_{\rm sc}$ and assuming Poisson fluctuations about the total quasiparticle number, GR noise will contribute $\sigma_{E\rm ,GR}^*=0.7\;\rm eV$ toward the inferred energy resolution. Thus, improvements to the resolution past this point will require reducing the background quasiparticle density (which would increase the quasiparticle lifetime $\tau_{\rm qp}\propto1/n_0$ \cite{pj2012generation} in systems dominated by quasiparticle recombination). Relatedly, a density of $n_0=25\;\mu\mathrm{m}^{-3}$ has been shown in the MKID literature \cite{de2011number} (equivalent to $\sigma_{E\rm ,GR}^*=0.15\;\rm eV$). This range of GR noise-dominated resolution is included in Fig.~\ref{fig:rqplot}. Further ideas to decrease $n_0$ focus on mitigating the effects of blackbody pair-breaking radiation \cite{chang2023, de2014fluctuations}. Increasing the quasiparticle lifetime has the added benefit of narrowing the bandwidth of $\tilde{s}$, improving the resolution by $1/\sqrt{\tau_{\rm qp}}$ \cite{golwala2000}.

\section{Application to a rare-event search and conclusion}

To explore the ramifications of a meV resolving KIPM detector, consider the case of searching for a light dark matter particle ($\chi$) via its elastic scattering on a heavier target (e.g. a silicon nucleus, $M$). The maximum energy transfer in such a two-body collision, if $m_{M} \gg m_{\chi}$, is $E_{\rm max}\approx 2 m_{\chi}^2 v_{\chi}^2 / m_M$, where $v_{\chi}$ is the particle velocity (capped at $\sim$600~km/s for dark matter due to Milky Way constraints). With an experimental threshold of $5\sigma_E$, a silicon target KIPM detector benefiting from SQL amplification with $\sigma_E \approx 200$~meV would be sensitive to a dark matter mass of $m_{\chi} \approx 55$~MeV. Current state-of-the art experiments probe dark matter masses of $m_{\chi}\sim100$~MeV \cite{angloher2023results, collar2018search}, which suggests significant expansion into previously unexplored dark matter coupling parameter space.

In summary, we have demonstrated the effect of low noise amplifications on KIPM detector readout. We ascertain that significant improvements to the substrate energy resolution on the order of $5\times$ can be achieved by pairing these devices with KI-TWPA parametric amplifiers. The limiting case of system noise quanta $N_{\rm sys}$ close to 1 can in part be achieved with better passive components and better RF engineering, making KI-TWPAs an ideal addition to next generation rare-event search experiments. However, resonator and quasiparticle physics, like the presence of two-level systems, can impact sensitivity and further work is required to overcome these limitations.

\begin{acknowledgments}
We thank N. Klimovich for assistance in understanding KI-TWPA behavior. We thank SLAC and FNAL collaborators for insightful conversations regarding MKIDs and parametric amplification. The research was carried out at the Jet Propulsion Laboratory, California Institute of Technology, under a contract with the National Aeronautics and Space Administration (80NM0018D0004). K.R. was supported in part by National Science Foundation PHY Grant 2209581. O.W. was supported by NASA NSTGRO Award 80NSSC20K1223. We also acknowledge support from the Department of Energy, DESC0011925F Fermilab, LDRD Subcontract 672112.
\end{acknowledgments}

\appendix

\section{KIPM detector non-idealities}\label{app:KIPMnonideal}

We have noticed that the transmission ($S21$) of a RF signal feedline can significantly deteriorate with increasing length and routing complexity. The deterioration is manifested by both large and broad $S21$ ripples and unintended resonances. The ripples can be larger than 10 dB and span several GHz, while the resonances can exhibit quality factors of $10^{2\text{--}3}$ and populate at a density of one per several hundred MHz \cite{chang2023}. We fabricated devices with identical feedlines with and without MKIDs. We observed that the broad ripples are independent of MKIDs, while the frequencies and quality factors of the feedline resonances appear to depend on the location of MKIDs resonances.

These transmission non-idealities make optimizing the readout power for MKIDs challenging. Not only does one need to adjust the readout power MKID-by-MKID, in some cases the $S21$ suppression can be more than 40 dB and prevent reaching the intended readout power at these frequencies. Second, MKIDs designed to exhibit the same resonant frequencies and quality factors can interact with the feedline resonances and result in substantially different responses. A typical distortion manifesting such interferences on MKIDs is a large ``coupling angle''. Ref.~\cite{khalil2012} proposed a semi-empirical model that considers a complex-number coupling quality factor, thus introducing an angle to that value, and helping accommodate rotated MKID resonance circles in the IQ plane away from the theoretical topology (i.e. tangent to and inside the feedline continuum circle). Ref.~\cite{khalil2012}'s model is applicable for small coupling angles, while for MKIDs that are affected by feedline resonances, their coupling angles appear to be as large as $\pi/2$. Finally, we also observed a general trend that large coupling angles appear more frequently on longer feedlines. We also suspect that such an interaction naturally promotes crosstalk coupling between MKIDs.

Disentangling competing effects is challenging, but we can likely attribute a significant portion of the above non-idealities to electromagnetic couplings between the device and holder (i.e. box-modes). We fabricated several feedlines (see Ref. \cite{chang2023} for device specifics) of the same design utilized in the main study with different routings on 3-inch wafers. The simplest routing connecting the two wirebonding ports by a straight line showed negligible rippling with no spurious resonances. For other devices with increasing routing complexities, such as more meandering, number of corners, and/or generally longer, the ripples grew and the resonances appeared more densely with, qualitatively, a positive correlation with the routing complexity. We tested these devices in a fully closed holder (nominal configuration), with holder lids removed, and with an Eccosorb (Laird Technologies) microwave absorber foam layer in the closed device holder. The Eccosorb strongly suppressed the feedline issue, leaving the $S21$ similar in appearance to a moderate-quality commercial coaxial cable that decreases toward higher frequencies by several dB in 1--10 GHz, no apparent ripple or resonance. For the lid-removed case, the $S21$ exhibited ripples, but the resonances disappeared due to the absences of a full enclosure. We perform similar experiments with $2\times2$ cm chips. The transmission spectrum was generally cleaner and exhibited fewer ripples and no resonance that could confound reconstructing the MKID data, regardless of the complexity of the routing or the device holder configuration \cite{wen2025strategic}. Such dependence on holder geometry and the presence/absence of absorptive layers suggests a dynamical coupling between device and environment that needs to be studied further.

\section{KI-TWPA non-idealities}\label{app:KITWPAnonideal}
%\subsection*{Lossy components}
%As discussed in the main text, this version of KI-TWPAs requires the installation of loss-inducing components, such as diplexers and isolators. These components are the main reason that these amplifiers do not reach the standard quantum limit in their current form. Fig.\;\ref{fig:improvement} Bottom illustrates how the incorporation of low-loss components would bring the system noise down to about 2 quanta. Improvement beyond this level would involve mitigation of on-chip losses that may occur in the dielectric or as a result of the non-linear four-wave mixing process. 

%\subsection*{Gain and noise ripples}
A significant non-ideality in this version of the KI-TWPA are the pronounced gain and noise ripples, both of which can directly degrade the SNR of resonator readout. A 4\,dB gain ripple is visible in Fig.\;\ref{fig:schematic} Bottom Right, and an approximately $\pm$40\% noise ripple is visible in Fig.\;\ref{fig:improvement} Bottom. Given the similar periodicity in the noise and gain ripples, we hypothesize that they arise from the same origin: impedance mismatches \cite{klimovich2022,klimovich2023}.

\begin{figure}
    \centering
    \includegraphics[width=.99\textwidth]{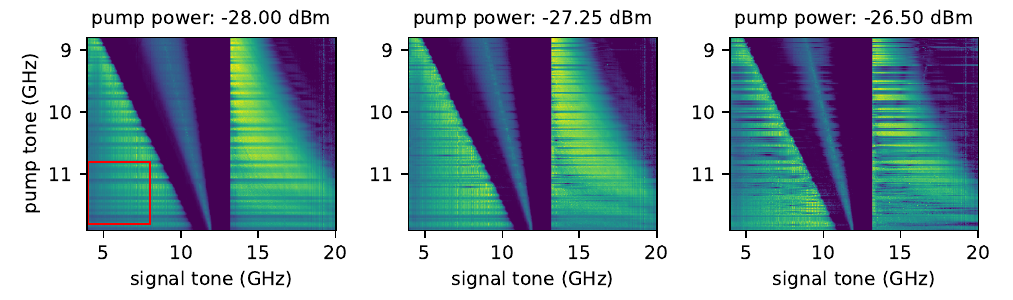}
    \includegraphics[width=.99\textwidth]{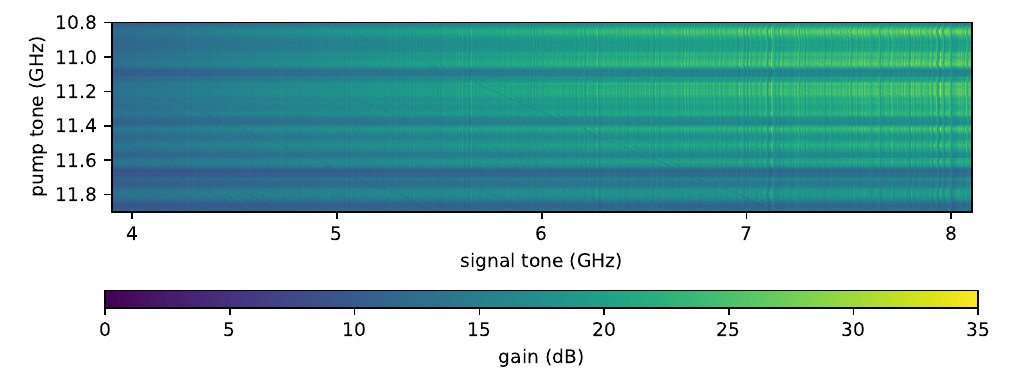}
    \caption{Four-wave mixing KI-TWPA gain versus pump frequency and pump power. \textit{Top:} gain performance at three different pump powers near the upper end of operable pump powers. The left band in each plot corresponds to the signal band, and the right band corresponds to the idler band. The band in the middle is centered on the pump frequency, which is visible as a thin, faint line. Recall the condition $\omega_s +\omega_i=2\omega_p$. The gain at the engineered bandgap from roughly 12 to 13\,GHz has been set to 0 in this plot, as the KI-TWPA has no transmission in this region. \textit{Bottom:} Zoom-in of the gain from 4 to 8\,GHz, for pump tones from 10.8 to 11.9\,GHz. The gain ripples are more visible, especially on the right side of the plot. There are also ripples that move with pump tone frequency.}
    \label{fig:KITWPA_gain}
\end{figure}

To better understand the origin and stability of these gain ripples, we investigated the performance of a similar KI-TWPA to the one presented in the main text, by varying the pump and signal tones. This particular device was engineered for four-wave mixing gain in the 4 to 8\,GHz range \cite{Faramarzi_2024}. The measurements shown in Fig.\;\ref{fig:KITWPA_gain} were performed at 4\,K in a system at the Goddard Space and Flight Center. In the top three panels of Fig.\;\ref{fig:KITWPA_gain}, we show the signal and idler bands of gain in the 4 to 20\,GHz region. The pump is faintly visible as a thin, bright line in the central band that spans from 9 to 12\,GHz. The bandgap is plotted as a hard cutoff from 12 to 13\,GHz, where this no transmission as a result of the engineered bandgap. The key point of top panels of Fig.\;\ref{fig:KITWPA_gain} is that the gain is very sensitive to pump frequency and pump power; As the pump power increases, there are specific pump frequencies that begin to show no gain, whereas there are other frequencies that maintain good gain over the full range of pump powers that are indicated.

The bottom panel of Fig.\;\ref{fig:KITWPA_gain} is a zoom-in of the red square from the top right panel. As noted in the top panels, the gain is dependent on the pump frequency: there are certain pump frequencies where the gain is poor. The zoom-in makes visible the individual gain ripples, most visible in the right section from 7 to 8\,GHz but also present in the entire 4 to 8\,GHz span. There are also ripples that move with pump frequency, shifting approximately two times that of a given shift in the pump tone (note the feature that moves from 6\,GHz to 7 GHz for pump tones from 11.3\,GHz to 11.8\,GHz. 

Thus, to achieve stable operation with maximal gain in the band of interest, one needs to carefully tune the pump frequency and power, performing a sweep similar to that presented above. It should be noted that even with the noise ripples in the main study (note that some of the MKIDs in Fig.\;\ref{fig:improvement} are located at noise ripple peaks), all of the resonators saw significant improvement in noise temperature and ultimately sensor resolution, as shown in Fig.\;\ref{fig:improvement}. The effect of the gain ripples is minimal so long as the KI-TWPA noise remains dominant over the HEMT noise (achievable with $\gtrapprox$ 15\,dB gain).
 
\bibliographystyle{JHEP}
\bibliography{refs.bib}

\end{document}